\documentclass[aps,prl,reprint,nobalancelastpage,nofootinbib]{revtex4-2}
\pdfoutput=1

\usepackage[utf8]{inputenc}

\usepackage{amsmath}
\usepackage{amssymb}
\usepackage[all]{xy}
\usepackage{tikz}
\usetikzlibrary{calc}
\usepackage{pgfplots}
\usepgfplotslibrary{fillbetween}
\usetikzlibrary{decorations.markings}

\usepackage[toc,page]{appendix}
\usepackage{enumerate}
\usepackage{tensor}
\usepackage{booktabs}
\usepackage{bbm}
\usepackage{youngtab}
\usepackage{slashed}
\usepackage{multirow}
\usepackage{makecell}
\usepackage{float}
\usepackage{comment}
\usepackage{verbatim}
\usepackage{xcolor}
\usepackage[framemethod=tikz]{mdframed}
\usepackage{soul}
\usepackage{mathtools}
\usepackage[inline]{enumitem}
\usepackage{alphabeta}
\usepackage{accents}
\usepackage{soul}

\usepackage{amsthm}

\newcommand{\ab}{|}
\newcommand{\der}{\partial}
\newcommand{\de}{\mathrm{d}}

\newcommand{\EFT}{{\mathrm{EFT}}}
\newcommand{\dS}{
{\mathrm{dS}}}
\newcommand{\s}{
{\mathrm{sp}}}
\newcommand{\UV}{
{\mathrm{UV}}}
\newcommand{\IR}{
{\mathrm{IR}}}

\newcommand{\e}{\mathrm{e}}
\newcommand{\I}{\mathrm{i}}

\newcommand{\p}{\mathrm{P}}

\allowdisplaybreaks

\usepackage{hyperref}
\hypersetup{
    colorlinks=true,
    linkcolor=cyan!45!blue,
    citecolor=cyan!45!blue,
    filecolor=cyan!45!blue,
    urlcolor=cyan!45!blue,
}

\makeatletter
\renewcommand{\p@subsection}{}
\renewcommand{\p@subsubsection}{}
\makeatother

\usepackage[backgroundcolor=white!95!red, linecolor=purple, bordercolor=purple, textsize=footnotesize]{todonotes}

\begin{document}
\numberwithin{equation}{section}

\title{Cosmological constraints from UV/IR mixing}

\author{Niccol\`{o} Cribiori}
\email{niccolo.cribiori@kuleuven.be}
\affiliation{Instituut voor Theoretische Fysica, KU Leuven, Celestijnenlaan 200D, B-3001 Leuven, Belgium}

\author{Flavio Tonioni}
\email{flavio.tonioni@unipd.it}
\affiliation{Dipartimento di Fisica e Astronomia, Università degli Studi di Padova, and INFN-Padova, via F. Marzolo 8, 35131 Padua, Italy}

\begin{abstract}
Holography and entropy bounds suggest that the ultraviolet (UV) and infrared (IR) cutoffs of gravitational effective theories are related to one another as a form of UV/IR mixing.
Motivated by this, we derive a bound on the allowed scalar field range in theories with cosmic horizons.
We show how this bound challenges several inflationary scenarios, such as $\alpha$-attractors and modular-invariant inflation.
Besides, we find a relation between the number of extra spatial dimensions and the tensor-to-scalar ratio.
\end{abstract}

\maketitle

\section{Introduction}

As of today, there is growing evidence that the ultraviolet (UV) and infrared (IR) regimes are not completely decoupled in quantum gravity.
In string theory this is certainly the case, as exemplified by modular invariance on the worldsheet; see \emph{e.g.} refs.~\cite{Kutasov:1990sv, Abel:2021tyt}.
From a bottom-up perspective, one may say that not all sets of IR parameters necessarily flow into a consistent UV completion.
This idea lies at the core of the Swampland Program \cite{Palti:2019pca, Agmon:2022thq}, and making it quantitative is a central challenge.
In this note, we investigate the consequences of the mixing between UV and IR regimes in models of phenomenological interest for cosmology, such as long-lived inflationary epochs.
We will motivate and investigate a specific manifestation of UV/IR mixing in the form of bounds on the UV and IR cutoff scales, $\Lambda_\UV$ and $\Lambda_\IR$, and we will argue that these bounds lead to constraints in large classes of models.

Our focus is on $d$-dimensional effective field theories (EFTs) in the presence of a cosmic horizon, such as models of cosmic acceleration with $d>2$.
The point we make is that, just like the requirement that $\Lambda_\IR \leq \Lambda_\UV$ induces an upper bound on the allowed inflaton field range \cite{Huang:2007qz,Hebecker:2018vxz, Scalisi:2018eaz}, holographic and entropy bounds \cite{Cohen:1998zx, Fischler:1998st, Bousso:1999xy, Bousso:2002ju} also imply the existence of a lower bound on the minimum field value.
This uniquely fixes a window where the inflationary dynamics should occur.
Concretely, we will argue that the inflaton $\varphi$ must take value within a finite interval, \emph{i.e.} $\varphi \in\; [\varphi_1, \varphi_2]$, and that the inflaton excursion during inflation, $\Delta \varphi$, must be such that
\begin{equation} \label{eq:deltaphirange}
    \kappa_d \Delta \varphi \leq  \dfrac{p}{2 \lambda} \, \ln \, \dfrac{(l_{\p, d}^d V_1)^{\frac{1}{1-(d-2)q}}}{(l_{\p, d}^d V_2)^{d-1}}.
\end{equation}
Here, $p$ and $q$ are parameters governing the amount of UV/IR mixing, $\lambda$ provides a measure of the decay rate of the UV cutoff with respect to $\varphi$, $V_{1,2}= V(\varphi_{1,2})$ the scalar potential evaluated at the endpoints of the allowed field range, and $\smash{l_{\p, d} = 1/m_{\p, d} = \kappa_d^{2/(d-2)}}$ is the Planck length.
The fact that general quantum gravity principles suggest that the parameters $p$, $q$ and $\lambda$ are order-one will be the seed of constraints on cosmological dynamics.

In several models of early cosmic inflation \cite{Starobinsky:1980te, Guth:1980zm, Linde:1981mu}, including $\alpha$-attractors \cite{Kallosh:2013hoa, Farakos:2013cqa, Ferrara:2013rsa, Kallosh:2013yoa, Garcia-Bellido:2014wfa, Galante:2014ifa, Roest:2015qya} and modular-invariant proposals \cite{Casas:2024jbw,Kallosh:2024ymt,Kallosh:2024pat}, we will argue that inflation does not stop as the boundary of the allowed inflaton range is reached.
A potential for early cosmic inflation needs to be such that the slow-roll parameters $\epsilon_V$ and $\ab \eta_V \ab$ are small during the inflationary phase.
We find that order-one parameters $p$, $q$ and $\lambda$ imply that the slow-roll parameters $\epsilon_V$ and $\ab \eta_V \ab$ are small all over the allowed range $\varphi \in \; [\varphi_1, \varphi_2]$, including its endpoints, where the EFT breaks down.
In other words, the point at which inflation ends lies outside the interval $[\varphi_1,\varphi_2]$.
As a consequence, one cannot describe the entire early cosmological dynamics -- eventually exiting the inflationary epoch -- within the given EFT.
From the relation between $\Delta \varphi$ and the number of e-folds $N_\e$, we can also get an upper bound on the latter.
We find that this bound is highly sensitive to the UV/IR mixing parameters.
For certain values of the parameters, the bound could forbid the minimum number of e-folds required to explain the CMB properties, $N_\e \gtrsim 60$ \cite{Baumann:2009ds}.

Our results align with other attempts to constrain EFTs of cosmic acceleration in a model-independent fashion, such as refs.~\cite{Huang:2007qz, Obied:2018sgi, Andriot:2018wzk, Garg:2018reu, Ooguri:2018wrx, Hebecker:2018vxz, Scalisi:2018eaz, Bedroya:2019snp, Seo:2021bpb, Rudelius:2021azq, Rudelius:2022gbz, Scalisi:2024jhq}, and we will comment on the relationship with them in detail.
More generally, it is believed that transplanckian scalar field excursions are pathological from an EFT point of view \cite{Ooguri:2006in}, at least for pure moduli.
Our bound in eq.~(\ref{eq:deltaphirange}) reflects this basic quantum gravity principle in theories with a potential.

It is also tempting to turn the logic around and use experimental inputs on inflationary observables, such as the number of e-folds $N_\e \gtrsim 60$ and the tensor-to-scalar ratio $r \lesssim 0.028$ \cite{Galloni:2022mok}, to constrain theoretical parameters.
In this way, we derive the bound
\begin{equation} \label{eq.:extradimensions}
    \sqrt{\dfrac{n}{n+2}} \leq \dfrac{1-3q}{1-2q} \, \dfrac{60 p}{N_\e^*} \, \sqrt{\dfrac{0.004}{r}} \, \ln \dfrac{10^8}{r}
\end{equation}
on the number $n$ of extra spatial dimensions.
By virtue of this relation, detection of primordial gravitational waves may constrain the number of extra dimensions.

The rest of the paper is organized as follows.
In sec.~\ref{sec.: UV/IR mixing and entropy bounds}, we review entropy bounds motivating our proposed UV/IR mixing relationship.
In sec.~\ref{sec.: bounds on inflationary dynamics}, we elaborate the implications of the latter on single-field inflationary models, discussing concrete examples in sec.~\ref{sec.: examples} and also in app.~\ref{app.: KKLT and LVS}.
A final discussion is in sec.~\ref{sec.: discussion}.

\section{Entropy bounds and UV/IR mixing} \label{sec.: UV/IR mixing and entropy bounds}

In this section, we review EFT constraints based on holography and entropy bounds and motivate the parameterization of UV/IR mixing that we employ throughout the note.
Related ideas are in refs.~\cite{Horvat:2010km, Ooguri:2018wrx, Castellano:2021mmx, Cribiori:2023ffn, Calderon-Infante:2023ler, Herraez:2024kux}.

A lightsheet of a $(d-2)$-dimensional surface is a $(d-1)$-dimensional surface generated by light rays that begin on the surface and extend orthogonally away from it with non-positive expansion.
The Covariant Entropy Bound (CEB) is the conjecture that the entropy $S$ on any lightsheet of a surface does not exceed the area $A_{d-2}$ of the surface in Planck units \cite{Bousso:1999xy, Bousso:2002ju}, \emph{i.e.}
\begin{equation} \label{CEB}
    S \leq \dfrac{1}{4} \, m_{\p, d}^{d-2} A_{d-2}.
\end{equation}
This bound will serve us as guiding principle to motivate a specific form of UV/IR mixing in accelerated expanding cosmologies.
In the rest of the section, we will ignore order-one coefficients, since we are only interested in estimating relative scalings.

Consider an EFT with $N$ light degrees of freedom.
If the EFT is spatially localized within a $(d-1)$-dimensional volume $\smash{\mathrm{vol}_{d-1}}$, given the UV-cutoff $\Lambda_\UV$, let us assume that the maximal EFT entropy scales as \cite{Cohen:1998zx, Banks:2019arz}
\begin{equation} \label{EFT entropy}
    S_\EFT = N \Lambda_\UV^{d-1} \mathrm{vol}_{d-1}.
\end{equation}
In particular, $S_\EFT$ is extensive, namely it scales with the volume and with the number of fields.
Then, one may get maximal entropy for maximal volume, hence the presence of the cutoff \cite{Castellano:2021mmx, Calderon-Infante:2023ler}.
At the same time, the total vacuum energy scales as
\begin{equation}
E = N \Lambda_\UV^d \mathrm{vol}_{d-1}.
\end{equation}
Importantly, we are working in a regime of weak gravitational coupling, which allows us to ignore gravitational interactions among the $N$ fields.

The system we have in mind is the observable universe in a spacetime with a cosmic horizon.
This is contained within a sphere of radius $R$ equal to the apparent cosmic horizon, \emph{i.e.} $\mathrm{vol}_{d-1}=R^{d-1}$ and $R = 1/H$, where $H$ is the Hubble parameter.
Examples of such spacetimes are cosmological solutions with accelerated expansion.
Because $1/H$ is the largest observable length, we set $\Lambda_\IR = H$.
To avoid gravitational collapse, we must require the Schwarzschild radius $\smash{r_{\mathrm{S}}(E) = (l_{\p, d}^{d-2} E)^{1/(d-3)}}$ not to be larger than $R$, giving
\begin{equation} \label{CKN BH stability}
    N \Bigl( \dfrac{\Lambda_\UV}{\Lambda_\IR} \Bigr)^d \leq \Bigl( \dfrac{m_{\p, d}}{\Lambda_\IR} \Bigr)^{d-2}
\end{equation}
and, correspondingly, $\smash{S_\EFT \leq N^{1/d} (m_{\p, d} R)^{(d-1)(d-2)/d}}$.
For $N=1$, we recover the Cohen-Kaplan-Nelson 
entropy bound \cite{Cohen:1998zx}.
The requirement in eq.~(\ref{CKN BH stability}) means that the largest admissible UV cutoff is such that\footnote{Although ref.~\cite{Ooguri:2018wrx} made considerations through the concept of a thermal bath, if we identify the temperature as $T = \Lambda_\UV$, then the results agree.}
\begin{equation}
\label{eq:LUVmax}
    (l_{\p, d} \Lambda_\UV)^d = \dfrac{(l_{\p, d} \Lambda_\IR)^2}{N}.
\end{equation}
Notice that the cutoff depends on $N$ and the size of the box $R=1/\Lambda_\IR$.
These quantities are not independent, else one could violate the CEB \emph{e.g}~by increasing $N$ arbitrarily at fixed $R$.

Having gained intuition on how entropy considerations lead to a link between UV and IR cutoff, we can look again at the CEB and rewrite eq.~(\ref{CEB}) as \cite{Herraez:2024kux}
\begin{equation} \label{EFT-CEB}
    (l_{\p, d} \, \Lambda_\UV)^{d-1} \leq l_{\p, d} \, \dfrac{\Lambda_\IR}{N},
\end{equation}
where we fixed $S = S_\EFT$. 
It remains to identify $\Lambda_\UV$.
This is conceptually harder than with $\Lambda_\IR$, for which a natural option is suggested by simple physical considerations.
On the other hand, several effects can concur in the determination of $\Lambda_\UV$ and certainly we cannot ignore gravity.
If we only look at gravitational interactions, an upper bound on the UV cutoff is given by the species scale $\smash{\Lambda_\s = m_{\p ,d}/N_\s^{1/(d-2)}}$ \cite{Veneziano:2001ah,Dvali:2007hz}. 
This is in fact an absolute upper bound on the UV cutoff of any $d$-dimensional EFT with $N_\s$ species; still, $\Lambda_\UV$ may be lowered by further (non-gravitational) effects.
Hence, at least we certainly have $\Lambda_\UV \leq \Lambda_\s$, as well as $N \leq N_\s$ by definition.
Note that the requirement $H \leq \Lambda_\UV$, in view of eq.~(\ref{eq:LUVmax}), implies the constraint $\smash{\Lambda_\UV \leq m_{\p, d}/N^{1/(d-2)}}$. 
With these facts in mind, given our ignorance on identifying unequivocally $\Lambda_\UV$, we take a conservative approach and assume the parameterization
\begin{equation} \label{LambdaUV-LambdaS}
    l_{\p, d} \Lambda_\UV = (l_{\p, d} \Lambda_\s)^{\frac{1}{(d-1) {p}}},
\end{equation}
where $p \leq 1/(d-1)$ is a positive constant. 
The highest value $p=1/(d-1)$ corresponds to taking $\Lambda_\UV = \Lambda_\IR$, if it is also $N=N_\s$, from eq.~(\ref{eq:LUVmax}).
This is then a limiting situation and gives the strongest constraints on cosmological models.\footnote{Notice that the fine details such as the order-one constants that we dropped (not being able to know them exactly) might alter this picture slightly, perhaps still allowing for $\Lambda_\IR \leq \Lambda_\UV$.}
Identifying a lower bound on $p$ is tricky. At least, we need $p \geq  \max \, \bigl[ \ln \, (l_{\p, d} \Lambda_\s) / \ln \, (l_{\p, d} \Lambda_\IR) \bigr]/(d-1)$ to ensure $\Lambda_\IR \leq \Lambda_\UV$.

The discussion so far supports the idea that the UV and IR cutoffs of an EFT cannot be chosen independently, which is a form of UV/IR mixing.
Although a precise relation between $\Lambda_\UV$ and $\Lambda_\IR$ may only be derived within a complete theory of quantum gravity -- which we lack --, the previous discussion eventually motivates a parameterization of UV/IR mixing such as
\begin{equation} \label{UV-IR mixing}
    l_{\p, d} \, \Lambda_\s \leq \dfrac{(l_{\p, d} \, H)^{p}}{N_\s^{q}},
\end{equation}
where $q$ is another non-negative parameter included for generality.\footnote{In general, we may write $\smash{N = N_\s^\alpha}$, for a power $\smash{\alpha \in\; [0,1]}$, since $\smash{1 \leq N \leq N_\s}$.
Hence, eqs.~(\ref{EFT-CEB}, \ref{LambdaUV-LambdaS}) can be combined into the inequality $\smash{l_{\p, d} \Lambda_\s \leq (l_{\p, d} H/N_\s^\alpha)^p}$, and we label $\smash{q = \alpha p}$.
So, for generality we include a second parameter $q$, but we do not change its relative sign to $p>0$.}
It must be bounded as $q \leq (1-p)/(d-2)$ to ensure the Hubble scale to never exceed the species scale.
For instance, taking $p=q=1/(d-1)$ reproduces eq.~(\ref{EFT-CEB}), with $\Lambda_\UV=\Lambda_\s$.
The purpose of this note is to study the consequence of eq.~(\ref{UV-IR mixing}) on accelerated expanding cosmologies.

We conclude with a comment on an alternative argument to evaluate the entropy stored in EFTs of cosmic inflation.
In ref.~\cite{Conlon:2012tz}, a central conclusion is that the maximal entropy of a four-dimensional EFT from modes at the horizon scales like $\smash{S_{\mathrm{h}} = N \Lambda_\UV^2/H^2}$, which dominates over the entropy in bulk modes.
Identifying $\smash{\Lambda_\UV = m_{\p, 4}/\sqrt{N}}$, this agrees with the CEB.
On the other hand, the entropy in bulk modes is estimated as $\smash{S_{\mathrm{b}} = \sqrt{N V}/H^2}$, with $V$ the potential.
Upon the same identification, we find $\smash{S_{\mathrm{b}} = m_{\p, 4}^2/\Lambda_\UV H}$, which differs from what we call $S_\EFT$.
Were it $\smash{S_\EFT \leq S_{\mathrm{b}}}$, we would get further support for the bound in eq.~(\ref{EFT-CEB}).
However, because $\smash{H \leq \Lambda_\UV \leq m_{\p, 4}}$, we cannot establish which of the inequalities $\smash{S_{\mathrm{b}} = (m_{\p, 4} H/\Lambda_\UV^2)^2 S_\EFT \lesseqgtr S_\EFT}$ holds true.
From $\smash{S_{\mathrm{b}} \leq S_{\mathrm{h}}}$, one may only say that $\Lambda_\IR \leq \Lambda_\UV$.

\section{Bounds on inflationary dynamics} \label{sec.: bounds on inflationary dynamics}

We focus on single-field models of cosmic inflation. 
In the slow-roll regime, the energy scale is set by the Hubble parameter $H$, which is related to the scalar potential $V$ as $\smash{H^2 = 2 l_{\p, d}^{d-2} V / [(d-1) (d-2)]}$.
By the very definition of UV cutoff we must have $\Lambda_\UV \geq \Lambda_\IR$, implying\footnote{Notice that one could impose directly $\Lambda_{\s}\geq \Lambda_\IR$, as in refs.~\cite{Hebecker:2018vxz,Scalisi:2018eaz}, but this would yield weaker bounds than those derived below.}
\begin{equation} \label{UV-cutoff}
    l_{\p, d} \Lambda_\s \geq (l_{\p, d} H)^{(d-1)p}.
\end{equation}
In accordance with general quantum gravity principles, we expect the species scale to decay exponentially at the boundary of field space \cite{Calderon-Infante:2023ler,vandeHeisteeg:2023ubh,vandeHeisteeg:2023uxj}, \emph{i.e.}
\begin{equation} \label{species scale}
    \Lambda_\s \geq m_{\p, d} \, \e^{- \kappa_d \lambda \varphi},
\end{equation}
for an order-one constant $\lambda$. 
Equivalently, the number of species $N_\s$ grows such that
\begin{equation} \label{species number}
    N_\s \leq N_0 \, \e^{\kappa_d (d-2) \lambda \varphi}.
\end{equation}
Possible polynomial dependencies in $N_0 = N_0(\varphi)$ are not going to be relevant for our analysis. At the boundary of field space, we expect these inequalities to be close to saturation.
From now on, we thus take $\smash{\Lambda_\s = m_{\p, d} \, \e^{- \kappa_d \lambda \varphi}}$ and $\smash{N_\s = N_0 \, \e^{\kappa_d (d-2) \lambda \varphi}}$.
Below, we explore the consequences of combining the bounds in eqs.~(\ref{UV-IR mixing}, \ref{UV-cutoff}).

First, a consequence of eqs.~(\ref{UV-cutoff}, \ref{species scale}) is the upper bound $\varphi \leq \varphi_2$, where
\begin{equation} \label{varphi - upper bound}
    \kappa_d \varphi_2 = - \dfrac{(d-1)p}{2 \lambda} \, \ln \, \bigl( l_{\p, d}^d V_2 \bigr).
\end{equation}
Second, a consequence of eqs.~(\ref{UV-IR mixing}, \ref{species scale}, \ref{species number}) is the lower bound $\varphi \geq \varphi_1$, where
\begin{equation} \label{varphi - lower bound}
    \kappa_d \varphi_1 = - \dfrac{1}{2 \lambda} \, \dfrac{p}{1 - (d-2) q} \, \ln \, \bigl( l_{\p, d}^d V_1 \bigr),
\end{equation}
where we highlight that $\smash{q \leq (1-p)/(d-2) < 1/(d-2)}$ because $p>0$, so $\varphi_2 \geq \varphi_1 \geq 0$.
Therefore, an inflationary theory fulfilling both eqs.~(\ref{UV-IR mixing}, \ref{UV-cutoff}) must evolve with the field $\varphi$ ranging at most through the interval 
\begin{equation} \label{eq:interval}
    \varphi_1 \leq  \varphi \leq \varphi_2.
\end{equation}
If a potential does not host an inflationary phase within the interval in eq.~(\ref{eq:interval}), the model does not comply with the proposed UV/IR-mixing constraints; see figs.~\ref{fig: field range} and~\ref{fig.: metastable dS - limiting case}.
In particular, a necessary condition on the inflationary dynamics is the inequality anticipated in eq.~(\ref{eq:deltaphirange}), \emph{i.e.}
\begin{equation} \label{varphi range}
     \Delta \varphi \leq \varphi_2 - \varphi_1 = \dfrac{p}{2 \kappa_d \lambda} \, \ln \, \dfrac{(l_{\p, d}^d V_1)^{\frac{1}{1-(d-2)q}}}{(l_{\p, d}^d V_2)^{d-1}},
\end{equation}
where $\smash{\Delta \varphi = \int_{t_\I}^{t_\e} \de t \, \sqrt{\dot{\varphi}^2(t)} \leq \varphi_2 - \varphi_1}$ is the field range spanned through the inflationary phase between the initial time $t_\I$ and the final time $t_\e$, assuming a monotonic evolution.
In eq. (\ref{varphi range}), we have a simple criterion that can be tested on any inflationary model more directly than the underlying principles in eqs. (\ref{UV-IR mixing}, \ref{UV-cutoff}).
If the interval $[\min \, (\varphi_\e, \varphi_\I), \max \, (\varphi_\e, \varphi_\I)]$ is not entirely contained within the interval $[\varphi_1, \varphi_2]$, then the inflationary dynamics extend beyond the region in which the EFT is well-defined.

\begin{figure}[ht]
\centering
    
\begin{tikzpicture}[xscale=1.15,yscale=4.0]

    \draw[->] (-0.5,0) -- (6.0,0) node[below]{$\varphi$};
    \draw[->] (0,-0.15) -- (0,0.92) node[left]{};
    \node[below left] at (0,0){$0$};

    \draw[densely dotted] (1.34,0) node[below]{$\varphi_1$} -- (1.34,0.63);
    \draw[densely dotted] (4.01,0) node[below]{$\varphi_2$} -- (4.01,0.11);

    \draw[domain=0.8:5.5, smooth, thick, variable=\x, orange] plot ({\x}, {1.5*exp(-0.65*\x)}) node[black,above]{$l_{\p, d} \Lambda_\s$};
    \draw[domain=0.8:4.8, smooth, thick, variable=\x, purple] plot ({\x}, {4/(\x+2)^2});
    \node[black,above] at (0.85,0.185){$(l_{\p, d} H)^{(d-1)p}$};
    \draw[domain=0.8:2.5, smooth, thick, variable=\x, green!80!black] plot ({\x}, {1.47/(\x+1)}) node[black, right] {$\dfrac{1}{N_\s^q} \, (l_{\p, d} H)^p$};

    \draw[ultra thick, cyan!80!black] (1.85,0) -- (3.5,0) node[pos=0.5, above, black]{inflation};

\end{tikzpicture}
    
\caption{A visualization of the field range singled out by eqs.~(\ref{UV-IR mixing}, \ref{UV-cutoff}). Due to UV/IR mixing, cosmic inflation should take place within the interval $\varphi \in \; [\varphi_1, \varphi_2]$.}
    
\label{fig: field range}
\end{figure}

We stress that the content of eq.~(\ref{varphi range}) is a global statement on the inflationary potential arising from UV/IR mixing.
% Importantly, although the value of the potential is constrained, the bound does not imply anything on the derivatives.
This means that, although the value of the potential is constrained locally, due to eqs.~(\ref{UV-IR mixing}, \ref{UV-cutoff}), we also have a global prediction based on eq.~(\ref{varphi range}), which is always very simple to test.
Yet, we are unable to draw local conclusions on the derivatives of the potential.
A priori, flat regions -- including mestastable de Sitter (dS) minima and maxima -- are allowed, as long as such flat regions are well within the interval in eq.~(\ref{eq:interval});
see fig.~\ref{fig.: metastable dS - limiting case}.
A key point in our analysis is that we expect all constants in eqs.~(\ref{varphi - upper bound}, \ref{varphi - lower bound}) to be order-one: the species-scale falloff rate is conjectured to be bounded as $\lambda \leq 1/\sqrt{d-2}$ \cite{vandeHeisteeg:2023ubh, Calderon-Infante:2023ler}, and we know that in the limiting case $p = q = 1/(d-1)$.
This expectation is also supported by comparison with other theorized general properties of UV-consistent EFTs \cite{Ooguri:2018wrx, Bedroya:2019snp, Rudelius:2021azq, vandeHeisteeg:2023uxj}, as we discuss below.

\begin{figure}[ht]
\centering
\begin{tikzpicture}[xscale=1.50,yscale=5.80,ppoint/.style={circle,minimum size=1.15mm,inner sep=0pt,outer sep=0pt,fill=purple,solid}]

    \draw[->] (-0.5,0) -- (4.8,0) node[below]{$\varphi$};
    \draw[->] (0,-0.10) -- (0,0.85) node[left]{};
    \node[below left] at (0,0){$0$};

    % the picture is for d=3, with p=1/3, q=1/6 and \lambda=1/2

    \draw[name path = ub, domain=0.32:4.25, smooth, thick, variable=\x, purple] plot ({\x}, {exp(-\x/2)});
    \node[purple] at (3.1,0.35){$(l_{\p,d} \Lambda_\s)^{\frac{2}{(d-1)p}}$};
    
    \draw[name path = lb, domain=0.32:4.25, smooth, thick, variable=\x, green!80!black] plot ({\x}, {exp(-\x)});
    \node[green!80!black] at (1.3,0.08) {$(N_\s^ql_{\p,d} \Lambda_\s)^{\frac{2}{p}}$};

    \draw[dotted] (0.32,0.73) -- (0.32,0) node[below]{$\varphi_1$};
    \draw[dotted] (4.25,0.12) -- (4.25,0) node[below]{$\varphi_2$};
    
    \tikzfillbetween[of = ub and lb]{orange, opacity=0.15};
    
    \draw[domain=0.32:4.25, smooth, thick, variable=\x, cyan] plot ({\x}, {0.0572*\x^2-0.4166*\x+0.8575});
    \node[above,cyan] at (3.3,0.10){$V$};

\end{tikzpicture}
    
\caption{Another visualization of the region (orange area) allowed by the bounds in eqs.~(\ref{UV-IR mixing}, \ref{UV-cutoff}). It might even allow for metastable dS minima (cyan curve).}
\label{fig.: metastable dS - limiting case}
\end{figure}

The bound described in eq.~(\ref{varphi range}) is novel, but
related to others in the literature.
The arguments for the refined dS conjecture in ref.~\cite{Ooguri:2018wrx} are based on the CEB.
Our results are compatible with it, but following a slightly different logic.
In this note, we are assuming eq.~(\ref{EFT-CEB}) -- parameterized in the form of eq.~(\ref{UV-IR mixing}) -- and we are also employing eq.~(\ref{UV-cutoff}), \emph{i.e.} $\smash{\Lambda_\IR \leq \Lambda_\UV}$: it is the combination of these two bounds that leads to eq.~(\ref{varphi range}).
In ref.~\cite{Ooguri:2018wrx}, only eq.~(\ref{EFT-CEB}) is made use of explicitly.\footnote{Of course, eq.~(\ref{UV-cutoff}), namely $\smash{\Lambda_\IR \leq \Lambda_\UV}$, also underlies ref.~\cite{Ooguri:2018wrx}, but it is not explicitly used to impose an additional constraint.}
Via the identification of the maximal UV-cutoff, the latter reads
\begin{equation} \label{EFT-CEB - refined dS}
    N^{\gamma} (m_{\p, d} R)^{\delta} \leq (m_{\p, d} R)^{\beta},
\end{equation}
with $\gamma = 1/d$ and $\delta = (d-1)(d-2)/d$ and $\beta = d-2$.
It is expected for the number of light states to grow like $N = N_0 \, \e^{b \kappa_d \varphi}$, and we stress that $b = (d-2) \lambda$ if we identify $N = N_\s$.
Via the relationship $\smash{l_{\p, d}^d V = 1/(m_{\p, d} R)^2}$, then, eq.~(\ref{EFT-CEB - refined dS}) implies
\begin{equation} \label{refined dS inequality - limiting case}
    l_{\p, d}^d V \geq N^{- \frac{2 \gamma}{\beta - \delta}} = \e^{- \frac{2 b \gamma}{\beta - \delta} \, \kappa_d \varphi}.
\end{equation}
The next step in ref.~\cite{Ooguri:2018wrx} -- which we have not taken here -- is to saturate the bound in eq.~(\ref{refined dS inequality - limiting case}) in view of the expectation that light degrees of freedom dominate the Hilbert space in the weak-coupling regime.
This eventually leads to the refined dS conjecture.
While we are not able to infer local predictions on the derivatives of the potential,
the refined dS conjecture infers the local bound \cite{Garg:2018reu,Ooguri:2018wrx}
\begin{equation} \label{refined dS}
\gamma_\dS \geq c \quad \lor \quad \eta_\dS \leq - c',
\end{equation}
for two positive universal order-one constants $c$ and $c'$, where in the single-field case $\smash{\gamma_\dS = \sqrt{(\der V/\der \varphi)^2} /( \kappa_d V)}$ and $\smash{\eta_\dS = (\der^2 V/ \der \varphi^2) / (\kappa_d^2 V)}$.
For a positive non-increasing potential, given any interval $\varphi \in \; [\varphi_a, \varphi_b]$, the condition $\smash{\gamma_\dS \geq c}$ implies 
\begin{equation} \label{global dS conjecture}
    V(\varphi_b) \leq V(\varphi_a) \, \e^{- c \kappa_d (\varphi_b - \varphi_a)}.
\end{equation}
Various arguments support the bound $c \geq 2/\sqrt{d-2}$ in the field-space asymptotics \cite{Bedroya:2019snp, Rudelius:2021azq, Shiu:2023nph}.
If we take $\varphi_{a,b} = \varphi_{1,2}$, this corresponds to taking $(d-1) p = \sqrt{d-2} \, \lambda$ and $q=1/(d-1)$ in eq.~(\ref{varphi range}).
Away from the field-space asymptotics, the transplanckian censorship conjecture \cite{Bedroya:2019snp} also leads to the inequality
\begin{equation} \label{TCC2}
    V(\varphi_b) \leq m_{\p, d}^d \, A \, \e^{- \frac{2 \kappa_d (\varphi_b - \varphi_a)}{\sqrt{(d-1)(d-2)}}},
\end{equation}
for an order-one constant $A$.
Demanding $\smash{H \leq \Lambda_\s}$ leads to the same conclusion, with $A<1$ \cite{vandeHeisteeg:2023uxj}; see also refs.~\cite{Scalisi:2018eaz, Bedroya:2019tba}.
Let $k$ be the constant such that $\smash{[l_{\p, d}^d V(\varphi_a)]^k = A}$.
Then, eq.~(\ref{varphi range}) would imply $\smash{\sqrt{d-1} \, p = \sqrt{d-2} \, \lambda}$ and $\smash{(d-1) q = (d-1-1/k)/(d-2)}$.

Let us now discuss some implications of eq.~(\ref{varphi range}).

\subsection{Limiting case} \label{sec:limiting case}

To begin with, let us discuss the limiting case, where we assume $\Lambda_\UV = \Lambda_\s$, \emph{i.e.}~we fix $p=q=1/(d-1)$.

According to eqs.~(\ref{UV-IR mixing}, \ref{UV-cutoff}), the potential is bounded as $\smash{l_{\p,d} \, \e^{-2 \lambda \varphi} \geq l_{\p, d}^{d-2} V \geq l_{\p,d} \, \e^{-2 \lambda \varphi}}$.
In other words, it is an exponential, with slope $\gamma = 2 \lambda$, namely
\begin{equation} \label{V - limiting case}
    V = \mu^d \, \e^{-2 \kappa_d \lambda \varphi}.
\end{equation}
%Our findings are consistent with ref.~\cite{Bedroya:2024zta}.
More generally, eq.~(\ref{V - limiting case}) is compatible  with general expectations from string compactifications, where the scalar potential may have a non-trivial profile in intermediate regions between the bulk and the field-space asymptotics, but is expected to vanish exponentially in limits such as zero string coupling and infinite volume \cite{Ooguri:2018wrx, Hebecker:2018vxz}.

\subsection{A bound on extra dimensions}

The UV/IR mixing parameters $p$ and $q$ and the species-scale decay rate $\lambda$ are central in our analysis.
On the other hand, we also have inflationary observables, such as the tensor-to-scalar ratio, $r$, the amplitude of scalar perturbations, $A_s$, the spectral tilt, $n_s$, and the number of e-folds, $N_\e$, whose values are constrained experimentally \cite{Planck:2018vyg, Planck:2018jri}.
In the spirit of ref.~\cite{Scalisi:2024jhq}, we may use observational data as inputs to constrain $p$, $q$ and $\lambda$, with $d=4$.

In slow-roll inflation, the power spectrum of curvature fluctuations for the mode wavenumber $k$ reads $\smash{\mathcal{P}_{\mathcal{R}}(k) = 2 \kappa_4^4 V/(3 \pi^2 r) \ab_{k = 1/l_{H}}}$ \cite{Baumann:2009ds}, with $r = 16 \epsilon_V$, where $l_{H}$ is the comoving Hubble radius.
Observationally, it is best-fit by the curve $\smash{\mathcal{P}_{\mathcal{R}}(k) = A_s \, (k/k_*)^{n_s-1}}$, where $A_s \simeq 2.10 \, \cdot 10^{-9}$ and $n_s \simeq 0.96$, for the pivot scale $k_* = 0.05 \, \mathrm{Mpc}^{-1}$ \cite{Planck:2018vyg, Planck:2018jri}.
By considering $V$ and $r$ to be approximately constant -- which means ignoring the $k$-dependence, as acceptable in a rough approximation being $n_s -1 \simeq - 0.04$ -- we may then rewrite eq.~(\ref{varphi range}) as
\begin{equation}
\label{eq:lambdabound1}
    \lambda \lesssim \dfrac{p}{\kappa_4 \Delta \varphi} \dfrac{1-3q}{1-2q} \, \ln \dfrac{10^8}{r}.
\end{equation}
This bound is different from the one of ref.~\cite{Scalisi:2024jhq} for two main reasons. First, ref.~\cite{Scalisi:2024jhq} uses $\smash{H_2 \simeq \kappa_4 \sqrt{V_2/3}}$, while our bound features the ratio $H_1/H_2$.
Second, even if we identify $\smash{H_1 \simeq 1/\kappa_4}$, our bound is more general due to the presence of the UV/IR mixing parameters $p$ and $q$. 
Then, one can calculate $\Delta \varphi$ as a function of $N_\e$, $r$ and possibly additional parameters of a given model, and substitute it in eq.~(\ref{eq:lambdabound1}).

Each specific inflationary model gives specific predictions on $\Delta \varphi$ and, in turn, constraints on $p,q$ and $\lambda$.
To remain as conservative as possible, we can replace $\Delta \varphi$ by the Lyth bound  \cite{Lyth:1996im, Boubekeur:2005zm, Boubekeur:2012xn}
\begin{equation}
    \kappa_4 \Delta \varphi \geq \dfrac{N_\e}{60} \sqrt{\dfrac{r}{0.002}},
\end{equation}
leading to 
\begin{equation}
    \lambda \leq  \dfrac{1-3q}{1-2q} \, \dfrac{60 p}{N_\e} \,  \sqrt{\dfrac{0.002}{r}} \, \ln \dfrac{10^8}{r}.
\end{equation}
This upper bound bound can further be combined with the lower bound \cite{Calderon-Infante:2023ler,vandeHeisteeg:2023ubh,vandeHeisteeg:2023uxj}
\begin{equation} \label{eq:lowerl}
    \lambda \geq \sqrt{\dfrac{n}{2(n+2)}}
\end{equation}
computed for $n$ extra spacetime dimensions, giving the constraint presented in eq.~(\ref{eq.:extradimensions}), \emph{i.e.}
\begin{equation} \label{eq:boundnextra}
    \sqrt{\dfrac{n}{n+2}} \leq \dfrac{1-3q}{1-2q} \, \dfrac{60 p}{N_\e} \, \sqrt{\dfrac{0.004}{r}} \, \ln \dfrac{10^8}{r}.
\end{equation}
This is a bound on the number $n$ of extra dimensions in terms of inflationary ($r, N_\e$) and UV/IR-mixing ($p, q$) parameters.
In particular, it implies that future detection of primordial gravitational waves might set an upper bound on the number of extra spacetime dimensions.
We also see that a smaller $r$ allows for a larger $n$.
For instance, if we saturate the experimental upper limit $r \lesssim 0.056$ and we take the indicative value $N_\e \simeq 60$ \cite{Planck:2018vyg}, we find $\sqrt{n/(n+2)} \lesssim 5.7 p \, (1-3q)/(1-2q)$.
One can see that if $q$ is close to the maximal value $q=(1-p)/2$, the bound becomes impossible to satisfy for any $n \geq 1$.
On the other hand, we may write $\smash{p \gtrsim (1/5.7) [(1-2q)/(1-3q)] \sqrt{n/(n+2)}}$.
For instance, exploiting the bound $1/3 \geq p$, we may infer $q \lesssim 0.29$ or $1/3 < q < 1/2$, computed for the least restrictive case $n=1$ (larger $n$ gives a lower maximal value).

It is apparent that the most critical dependence is the one on $r$: finding experimental evidence for a very small $r$ would relax the restrictions on $n$.

If the assumption on the constant potential is dropped, we can still write
\begin{equation} \label{eq:Nbound}
    \dfrac{N_\e}{60} \sqrt{\dfrac{r}{0.002}} \leq \dfrac{p}{2 \lambda} \, \ln \, \dfrac{(\kappa_4^4 V_1)^{\frac{1}{1-2q}}}{(\kappa_4^4 V_2)^3}.
\end{equation}
One can then repeat a similar analysis as above.

\section{Examples} \label{sec.: examples}

In this section we overview the implications of eqs.~(\ref{varphi - upper bound}, \ref{varphi - lower bound}, \ref{varphi range}) for a number of proposals of slow-roll cosmic inflation \cite{Martin:2013tda}.
In app. \ref{app.: KKLT and LVS}, we discuss proposed realizations of string-theoretic dS vacua.

We have bounds constraining the scalar potential, namely in eqs. (\ref{UV-IR mixing}, \ref{UV-cutoff}).
As discussed, a (weaker) global statement is the bound in eq.~(\ref{varphi range}), relating the largest admissible field excursion to the global variation of the scalar potential.
This results in a simple criterion that we can  apply to readily raise tensions in given inflationary models.

\subsection{Chaotic inflation} \label{ssec.: chaotic inflation}

Let us consider a prototypical chaotic inflationary model with a power-law potential \cite{Linde:1983gd}
\begin{equation} \label{power-law potential: V}
    V(\varphi) = m_{\p, d}^d \, (\beta \kappa_d \varphi)^{2m},
\end{equation}
where $m, \beta>0$ are parameters.
Such models are in tension with data \cite{BICEP:2021xfz}, but it is nevertheless instructive for us to discuss them.
Indeed, the problem we will stumble upon will be recurrent in other models, too, that are instead compatible with current observations.
This problem is the fact that $\varphi_e$ lies outside the allowed inflaton field range in eq.~(\ref{eq:interval}).
The advantage of the present setup is that we will be able to work at each step with analytic expressions.

The values of $\varphi_{1,2}$ setting the boundaries of the allowed inflaton range (\ref{eq:interval}) are the solutions to the equations
\begin{equation}
    \e^{-h_{1,2} \lambda \kappa_d \varphi_{1,2}} = (\beta \kappa_d \varphi_{1,2})^{m},
\end{equation}
where, for brevity, we defined $h_1=[1-(d-2)q]/p$ and $h_2=1/[(d-1)p]$.
Notice that their ratio is bounded as $h_1/h_2 = [1-(d-2) q] (d-1) \leq d-1$.
We also define $\xi_{1,2}=h_{1,2} \lambda$.
An analytic solution can be written in terms of the principal branch of the Lambert $W$-function, \emph{i.e.}
\begin{equation} \label{power-law potential: solutions}
    \kappa_d \varphi_{1,2} = \dfrac{m}{\xi_{1,2}} \, W_0 \biggl( \dfrac{\xi_{1,2}}{m \beta} \biggr).
\end{equation}
Below, we argue that, for generic order-one parameters $\smash{\xi_{1,2}}$, the point where inflation would stop lies outside the allowed inflaton range.
For brevity, we fix $d=4$.

Inflation can occur as long as the slow-roll parameters $\epsilon_V$ and $\smash{\ab \eta_V \ab}$ are small.
Since in this class of models $\smash{\epsilon_V = 2m^2/(\kappa_4 \varphi)^2}$ and $\smash{\eta_V = (2-1/m) \, \epsilon_V}$ we focus on the condition $\epsilon_V \ll 1$ for brevity.
In particular, inflation stops at the latest when $\kappa_4 \varphi \simeq \sqrt{2} \, m$. By plugging in the solution for $\varphi_{1,2}$, we find
\begin{equation} \label{eq:eVchaotic}
     \sqrt{\epsilon_V(\varphi_{1,2})} = \frac{\sqrt{2} \, \xi_{1,2}}{W_0 \Bigl( \dfrac{\xi_{1,2}}{m \beta} \Bigr)}.
\end{equation}
To have $H \ll m_{\p,d}$, one must fix $\beta \ll 1$. Hence, if we assume that $\xi_{1,2} = O(1)$, we may conclude that\footnote{At large $x$, we have $W_0(x) = \ln x - \ln \, \ln x + O[(\ln \ln x)/ \ln x]$ \cite{deBruijn}.}
\begin{equation}
    \sqrt{\epsilon_V(\varphi_{1,2})} \simeq \frac{\sqrt{2} \, \xi_{1,2}}{\ln \Bigl( \dfrac{\xi_{1,2}}{m \beta} \Bigr)} \ll 1.
\end{equation}
Hence, as the inflaton rolls towards the origin, it will cross the value $\smash{\varphi = \varphi_1}$ while $\epsilon_V$ is still tiny; in other words, one would have $\varphi_\e < \varphi_1$.
If instead $\xi_{1,2} \ll m \beta$, eq.~(\ref{eq:eVchaotic}) behaves as $\smash{\sqrt{\epsilon_V(\varphi_{1,2})} \simeq \sqrt{2} m \beta \ll 1}$.
Because the function $W_0$ is monotonically increasing, we thus learn that the only way to achieve $\smash{\sqrt{\epsilon_V(\varphi_1)} > 1}$ is if $h_{1,2} \gg O(1)$.
Otherwise, we conclude that at $\varphi_{1,2}$, where the EFT violates UV/IR-mixing, inflation is still dynamically allowed since $\epsilon_V$ and $\eta_V$ are small.

Let us give a concrete example.
An inflationary scale $H \lesssim 10^{-5}m_{\p,4}$ \cite{Planck:2018jri} requires $\smash{\beta \kappa_4 \varphi \lesssim 10^{-5/m}}$, so we fix $\smash{\beta = 10^{-4/m}}$.
To be optimistic and get a large range $\varphi_2 - \varphi_1$, we pick the smallest $h_2$ and then the largest $h_1$.
This means first picking $p = p_{\mathrm{max}}=1/3$, \emph{i.e.} $h_2=1$, and then $q = q_{\mathrm{min}} = 0$, \emph{i.e.} $h_1 = 3$.\footnote{Of course, maximizing the range $\smash{\varphi_2 - \varphi_1}$ according to eq.~(\ref{power-law potential: solutions}) is a complicated optimization problem.
We stick to the choice in the main text since the example is just for illustrative purposes.}
The largest $\epsilon_V$-value within the interval in eq.~(\ref{eq:interval}) is $\epsilon_V(\varphi_1)$.
If \emph{e.g.} $\lambda = 1/\sqrt{2}$, we see that $\epsilon_V(\varphi_1) > 1$ only for $m \geq 3$.
In that case, however, $\epsilon(\varphi_2)$ is too large for realistic inflationary scenarios; \emph{e.g.} for $m=3$, $\epsilon_V(\varphi_1) = 2.20$ and $\epsilon_V(\varphi_2) = 0.56$. Increasing $m$ will increase $\epsilon_V$ at both points.

\subsection{\texorpdfstring{$\boldsymbol{\alpha}$}{alpha}-attractors}

A well-studied class of four-dimensional inflationary models leading to testable predictions collectively go by the name of $\alpha$-attractors
\cite{Kallosh:2013hoa,Farakos:2013cqa,Ferrara:2013rsa,Kallosh:2013yoa,Galante:2014ifa}. For recent reviews, see refs.~\cite{Antoniadis:2024hvw, Kallosh:2025jsb, Kallosh:2025ijd}. 
Two basic classes of $\alpha$-attractors are the so called T-models and E-models, with respective potentials
\begin{subequations}
\begin{align}
    V_{\mathrm{T}}(\varphi) & = \Bigl( \dfrac{\mu}{\kappa_4} \Bigr)^4 \, \tanh^{2m}\dfrac{\kappa_4\varphi}{\sqrt{6\alpha}}, \label{alpha-attractor potential T} \\
    V_{\mathrm{E}}(\varphi) & = \Bigl( \dfrac{\mu}{\kappa_4} \Bigr)^4 \, \bigl(1 - \, \e^{-\sqrt{\frac{2}{3 \alpha}} \, \kappa_4 \varphi}\bigr)^{2m}, \label{alpha-attractor potential E}
\end{align}
\end{subequations}
where $\alpha, \mu, m > 0$ are parameters.
For $\alpha \gg 1$, both models reduce to chaotic inflation and our previous discussion applies.
Instead, the attractor behavior is manifest when $\alpha$ is order-one, such that the spectral index $n_s$ and tensor-to-scalar ratio $r$ only depend on $\alpha$ and the number of e-folds $N_\e$.
In particular, for $\alpha=m=1$, the E-model reduces to the Starobinsky model \cite{Starobinsky:1980te}.\footnote{In supergravity embeddings, $\alpha$ is related to the field-space curvature through $\kappa_4^2 R_\mathrm{M} = - 2/3\alpha$. However, especially when non-linearly realized, supergravity alone does not guarantee UV-consistency.
Indeed, embedding $\alpha$-attractors into a microscopic theory of quantum gravity, such as string theory, remains an open challenge.
We refer to refs.~\cite{Lust:2023zql, Brinkmann:2023eph, Antoniadis:2024ypf} for recent discussions on the specific case of the Starobinsky model.}

By following the same strategy as for chaotic inflation, we will show that for $\alpha$-attractor models, too, the inflaton range allowed by eq.~(\ref{eq:interval}) is such that the EFT loses validity while inflation is still happening.
In other words, $\varphi_\e$ is not contained in the allowed interval in eq.~(\ref{eq:interval}).
Again, this will occur generically for order-one parameters, but we cannot exclude the possibility that by appropriately fine-tuning
the parameters $p$, $q$ and $\lambda$ one can avoid our conclusion.

The starting point is to determine the extrema, $\varphi_{1,2}$ of the interval in eq.~(\ref{eq:interval}), \emph{i.e.} the solutions to
\begin{subequations}
\begin{align}
    \e^{-2 h_{1,2} \lambda \kappa_4 \varphi_{1,2}} = V_{\mathrm{T}}(\varphi_{1,2}), \label{eq:varphi12cond_1} \\
    \e^{-2 h_{1,2} \lambda \kappa_4 \varphi_{1,2}} = V_{\mathrm{E}}(\varphi_{1,2}), \label{eq:varphi12cond_2}
\end{align}
\end{subequations}
again for $h_1=[1-(d-2)q]/p$ and $h_2=1/[(d-1)p]$.
We are unable to find analytic solutions, but the problem can be understood qualitatively and discussed numerically.

To start, we recall that we must fix $\mu \ll 1$ in order to have $H \ll m_{\p,4}$.
Then, given $\xi_{1,2} = h_{1,2} \lambda$, we identify three possible situations; see also fig. \ref{fig.: alpha-attractors vs. CEB}.
\begin{enumerate}[label=(\alph*)]
    \item If $\xi_2 \leq \xi_1 \leq O(1)$, the equations have solutions only away from the origin, in the plateau-like region. \label{case a}
    \item If $\xi_2 \leq O(1) \leq \xi_1$, $\varphi_1$ is in the power-law region and $\varphi_2$ still in the plateau. \label{case b}
    \item If $O(1) \leq \xi_2 \leq \xi_1$, the solutions take place in the power-law region near the origin. \label{case c}
\end{enumerate}
For case \ref{case c}, we refer directly to ssec.~\ref{ssec.: chaotic inflation}.
Below, we thus discuss only cases \ref{case a} and \ref{case b}.
We remark that we are considering $q < p$; if $p=q$, we already know from ssec.~\ref{sec:limiting case} that the potentials would not be compatible with the bound since they are monotonically-increasing for $\varphi > 0$.

\begin{figure}[ht]
\centering
\begin{tikzpicture}[xscale=0.45,yscale=4.25,ppoint/.style={circle,minimum size=1.15mm,inner sep=0pt,outer sep=0pt,fill=purple,solid}]

    \draw[->] (-1.5,0) -- (15.5,0) node[below]{$\varphi$};
    \draw[->] (0,-0.10) -- (0,0.8) node[left]{};
    \node[below left] at (0,0){$0$};
    
    \draw[domain=1.85:14.65, smooth, thick, variable=\x, magenta] plot ({\x}, {exp(-0.106*\x)}) node[black,below]{$\e^{- 2 \xi_2 \kappa_4 \varphi}$};
    \draw[domain=1.0:8.5, smooth, thick, variable=\x, magenta] plot ({\x}, {exp(-0.35*\x)}) node[black,right]{$\e^{- 2 \xi_1' \kappa_4 \varphi}$};
    \draw[domain=0.25:6.0, smooth, thick, variable=\x, magenta] plot ({\x}, {exp(-2*\x)}) node[black,below right]{$\e^{- 2 \xi_1'' \kappa_4 \varphi}$};

    \draw[domain=-0.8:15.0, smooth, thick, variable=\x, cyan!85!blue] plot ({\x}, {(1-2*exp(-\x)+1.1*exp(-2*\x))/4}) node[black,above,align=center]{$\kappa_4^4 V_{\mathrm{E}}$};

    \draw[densely dotted] (4.06,0) node[black, below]{$\varphi_1'$} -- (4.06,0.24) node[ppoint]{};
    \draw[densely dotted] (1.10,0) node[black, below]{$\varphi_1''$} -- (1.10,0.11) node[ppoint]{};
    \draw[densely dotted] (13.0,0) node[black, below]{$\varphi_2$} -- (13.0,0.25) node[ppoint]{};

\end{tikzpicture}
    
\caption{A visualization of the solutions for eqs.~(\ref{varphi - upper bound}, \ref{varphi - lower bound}) for the E-model potential: for fixed $\varphi_2$, larger values $\xi_1'' > \xi_1'$ provide more room for inflation, being $\varphi_1'' < \varphi_1'$.
The plot for the T-model is qualitatively similar for $\varphi>0$.}
\label{fig.: alpha-attractors vs. CEB}
\end{figure}

In case \ref{case a}, both $\varphi_1$ and $\varphi_2$ are in the plateau region, therefore the solution does not stop inflating when $\varphi$ crosses the point $\varphi_1$, where the potential is still shallow.
Hence, $\varphi_\e<\varphi_1$ in this case.
Yet, one can ask whether the number of e-folds between $\varphi_2$ and $\varphi_1$, $N_{21}$, is very large.
For both classes of $\alpha$-attractors, at leading order in $N_{21}$, we can calculate
\begin{equation} \label{eq:N21alphattr}
    \kappa_4 \varphi_2 \simeq \sqrt{\dfrac{3 \alpha}{2}} \, \ln N_{21},
\end{equation}
which depends only on $\alpha$, but not on $m$ and $\mu$. By plugging in eqs.~(\ref{eq:varphi12cond_1}, \ref{eq:varphi12cond_2}), we find
\begin{equation}
    N_{21} \simeq \bigl[\kappa_4^4 V_{\mathrm{T}, \mathrm{E}} (\varphi_2)]^{- \frac{1}{\sqrt{6 \alpha}} \frac{1}{\xi_2}}.
\end{equation}
Approximating $\smash{V_{\mathrm{T}, \mathrm{E}} \simeq (\mu/\kappa_4)^4}$, we get 
\begin{equation}
    \smash{N_{21} \simeq \mu^{- \sqrt{\frac{8}{3 \alpha}} \frac{1}{\xi_2}}},
\end{equation}
which is extremely sensitive to the values $p$ and $\lambda$ in $\xi_2 = \lambda/3p$.
For instance, let $\mu = 10^{-5/2}$, $\alpha=1$, and $\lambda=1/\sqrt{2}$: for the limiting case $p=1/3$ we find $N_{21} \simeq 6 \cdot 10^6$, while for $p=1/12$ we get $N_{21} \simeq 28$.
By keeping the first correction to eq.~(\ref{eq:N21alphattr}), too, we get
\begin{align}
    \ln N_{21} \simeq \ln \, \bigl( \e^{\sqrt{\frac{2}{3 \alpha}} \, \kappa_4 \varphi_2} - \e^{\sqrt{\frac{2}{3 \alpha}} \, \kappa_4 \varphi_1} \bigr).
\end{align}
We write $q = (1-p)/2 - \delta/2$, with $\delta>0$ parameterizing how far we are from saturating the upper bound on $q$.
Then, we can write $\smash{\e^{\kappa_4 \varphi_1} = \e^{\kappa_4 \varphi_2/[3(p+\delta)]}}$, whence
\begin{equation}
    \ln N_{21} \simeq \sqrt{\dfrac{2}{3 \alpha}} \, \kappa_4 \varphi_2 + \ln \, \bigl( 1 - \e^{\sqrt{\frac{2}{3 \alpha}} \, \frac{1 - 3(p+\delta)}{3(p+\delta)} \, \kappa_4 \varphi_2} \bigr).
\end{equation}
We see that we get a negative contribution from the logarithm in the right-hand side -- which is bigger in absolute value for smaller $p+\delta$ --, making $N_{21}$ smaller.

In case \ref{case b}, the value $\varphi_1$ is such that $\epsilon_V$ may be large.
Yet, $N_\e$ is dominated by the value of $\varphi_2$, according to the discussion of case \ref{case a}.
This shows that, if $p$ and $q$ are not close to their limiting values, the bound of eq~(\ref{varphi range}) is not extremely restrictive in this case either.

We now illustrate a numerical example. 
Again, we fix $(h_1,h_2)=(3,1)$ and $\lambda = 1/\sqrt 2$.
For an inflationary scale $H \simeq 10^{-5} m_{\p,4}$, we pick $\mu \simeq 10^{-5/2}$.
We choose $\alpha=m=1$ because it corresponds to the Starobinsky model for the E-potential, but we checked that for values $(\alpha, m) \in \; [10^{-1},10^2]$ our considerations are qualitatively similar. 
The solutions to eqs.~(\ref{eq:varphi12cond_1}, \ref{eq:varphi12cond_2}) are $\kappa_4 (\varphi_1, \varphi_2) \simeq (5.44, 16.28)$ for the T-model and $\kappa_4 (\varphi_1, \varphi_2) \simeq (5.43, 16.28)$ for the E-model. 
At $\varphi_2$ the value of the slow-roll parameters is $(\epsilon_V, |\eta_V|) \simeq (1.51 \cdot 10^{-11}, 4.50 \cdot 10^{-6})$ for the T-model and $(\epsilon_V, |\eta_V|) \simeq (3.79 \cdot 10^{-12}, 2.25 \cdot 10^{-6})$ for the E-model. 
At $\varphi_1$ the value of the slow-roll parameters is $(\epsilon_V, |\eta_V|) \simeq (7.40 \cdot 10^{-4}, 3.00 \cdot 10^{-2})$ for the T-model and $(\epsilon_V, |\eta_V|) \simeq (1.92 \cdot 10^{-4}, 1.58 \cdot 10^{-2})$ for the E-model.
Hence, we see that the inflationary EFT stops being valid while inflation is still dynamically allowed. 
Besides, we find that $N_{21} \simeq 6980$ and $N_{21} \simeq 7037$ for the T- and E-models, respectively.
However, even small variations on $p$ and $q$ have a drastic impact on $N_{21}$.
For instance, let us pick the maximal value $p=1/3$ and $q$ close to the maximal value ($q=1/3)$, \emph{e.g.} $q=0.3$.
This means $h_1=1.2$ and $h_2=1$, deviating from the optimistic case above.
We find $\kappa_4 (\varphi_1, \varphi_2) \simeq (13.57, 16.28)$ for both the T- and the E-model.
The values of the slow-roll parameters are still tiny across the entire interval.
However, in this case we see that $N_{21} \simeq 9.14$ for both models.

\subsection{Modular-invariant cosmologies}

Recently, a novel class of inflationary models has been proposed in which the action of the scalar field is required to be modular-invariant \cite{Casas:2024jbw, Kallosh:2024ymt, Kallosh:2024pat,Aoki:2024ixq, Carrasco:2025rud}.
In this class, one can construct potentials that at large field value resemble closely the $\alpha$-attractor shape, but that may differ at small field values.
By direct inspection, one can check that several potentials constructed in this way are of the generic type 
\begin{equation}
    V = \Bigl( \dfrac{\mu}{\kappa_4} \Bigr)^4 \, (1 - \rho \, \e^{-a \kappa_d \varphi})^m,
\end{equation}
where $\rho \leq 1$ and $a$ is order-one. In particular, this is the case for the explicit models of refs.~\cite{Casas:2024jbw, Kallosh:2024ymt}.
Then, for these potentials, our previous analysis on standard $\alpha$-attractors can be repeated almost without modifications and our conclusions are unchanged.

\section{Discussion} \label{sec.: discussion}

Motivated by general ideas in quantum gravity, such as holography and entropy bounds, we have investigated a specific form of mixing between UV and IR cutoff scales in EFTs describing cosmic acceleration.
This resulted in bounds constraining the scalar potential, in eqs. (\ref{UV-IR mixing}, \ref{UV-cutoff}), and in the bound in eq.~(\ref{varphi range}), relating the largest admissible field excursion to the global variation of the scalar potential, which represents a simple criterion.
For models of early cosmic inflation, we argued that the bound is challenging to obey while describing the whole inflationary epoch within a given EFT.
Although with a number of ways out, which we discussed, our results thus hint at a generic tension between four-dimensional inflationary models and general quantum gravity principles.
This follows despite more conservative assumptions than earlier studies \cite{Ooguri:2018wrx, Bedroya:2019snp, Agrawal:2018own, Agrawal:2018rcg}.
The tension emerges when combining large field excursions with comparatively small variations in the potential, as supposedly happens in plateau-like regions. 
On the other hand, because the dark energy-dominated era has only lasted for less than one e-fold, it is harder to formulate constraints for the present universe that are based on a bound on the total admissible field displacement, like ours.
Nevertheless, constraining models of quintessence with large field excursions may still elucidate the fate of cosmic horizons in string compactifications.

We also note that the assumption of slow-roll can be relaxed.
There exist general classes of potentials -- expected in string compactifications, like multi-exponential potentials \cite{Hebecker:2018vxz, VanRiet:2023cca} -- such that the dynamics is effectively single field-like and that the kinetic and potential energies evolve at a fixed non-negligible ratio \cite{Shiu:2023fhb}.
This fact is crucial in extensions of the distance conjecture in the presence of a potential \cite{Mohseni:2024njl, Debusschere:2024rmi} and, likewise, it immediately extends the applicability of eq.~(\ref{varphi range}).
Since $H^2 = \sigma \kappa_d^2 V$ beyond the slow-roll approximation, for an order-one constant $\sigma$, our whole analysis still carries out.

The specific implications of eq.~(\ref{varphi range}) depend on the precise values of $p$, $q$ and $\lambda$, which we have refrained from fixing, but we can list some general observations.
\begin{itemize}[leftmargin=*]
    \item As showed in examples, eq.~(\ref{varphi range}) can lead to a situation in which inflation stops neither at $\varphi_1$ nor at $\varphi_2$ because the potential is still sufficiently flat at those points.
    This happens generically for order-one parameters $p$, $q$ and $\lambda$.
    In this case, we are not allowed to describe the whole inflationary dynamics while fulfilling the proposed bounds on UV/IR mixing.
    \item Another constraint stemming from eq.~(\ref{varphi range}) is in the form of an upper bound on the number of e-folds.
    For certain models, such as \emph{e.g.} $\alpha$-attractors, this bound is highly sensitive to the values of $p$, $q$ and $\lambda$.
    \item An important caveat is that, during reheating, the CEB and entropy bounds in general are not well-motivated due to the lack of a cosmic horizon.
    However, if the potential is shallow across the whole interval $[\varphi_1, \varphi_2]$, as is actually happening in the examples, the EFT becomes invalid while still in the proper inflationary epoch, before the reheating phase.
    \item A general limitation of the bound in eq.~(\ref{varphi range}) is that it requires knowledge of the potential at $\varphi_{1,2}$, but the inflationary phase may take place well in the interior of the interval in eq.~(\ref{eq:interval}).
    \item As a conclusive comment, which applies to all setups considered in this work, the fact that $\kappa_4\varphi_{1,2}$ are large but not parametrically so suggests that these models do not strictly live at the boundary of field space.
    In this sense, UV/IR mixing is teaching us that microscopically consistent models of inflation, if they exist, cannot be in asymptotic regions of field space, where one typically gains computational control.
    Since computational control in the bulk is usually a consequence of supersymmetry, it is not clear to us how it can be gained in these inflationary models; see \emph{e.g.}~refs.~\cite{Cicoli:2021skd, Cicoli:2021fsd}.
\end{itemize}

There are several future directions that one could pursue.
The most natural one consists in sharpening our knowledge of the parameters $p$ and $q$ in eq.~(\ref{UV-IR mixing}), which will provide much more precise predictions on the allowed potential variation across the maximum field range, as well as on the allowed number of e-folds. 
Another natural one consists in generalizing our analysis to multi-field models.
The current conclusions technically hold only for EFTs that are effectively one-dimensional in field space during the inflationary phase.
Although this is a feature of a large class of cosmological models, alternatives exist and conclusions might be different.

\begin{acknowledgments}
\subsection*{Acknowledgments}
We thank F.~Farakos, A.~Paraskevopoulou, M.~Scalisi, and T.~Van Riet for helpful discussions.
The work of NC is supported by the Research Foundation Flanders (FWO grant 1259125N).
The work of FT is funded by the European Union - NextGenerationEU/PNRR mission 4.1; CUP: C93C24004950006.
\end{acknowledgments}

\appendix

\section{KKLT and LVS scenarios} \label{app.: KKLT and LVS}

In this appendix we apply qualitatively our bounds on two concrete proposals for constructing dS vacua in string theory; for general reviews, see \emph{e.g.} refs.~\cite{Danielsson:2018ztv, Cicoli:2018kdo, Cicoli:2023opf}.

\subsection{KKLT}

The Kachru-Kallosh-Linde-Trivedi (KKLT) scenario \cite{Kachru:2003aw} is a proposal for constructing dS vacua in string theory.
As a four-dimensional supergravity construction, it features a potential generated by 3-form fluxes in a type-IIB Calabi-Yau orientifold, together with non-perturbative corrections and an uplift term associated to a supersymmetry-breaking anti-D3-brane.

A complete understanding of the string embedding of this model is still elusive.\footnote{See refs.~\cite{Gao:2020xqh, Lust:2022lfc, Blumenhagen:2022dbo} for recent criticisms, and ref.~\cite{McAllister:2024lnt} for recent efforts towards complete top-down realizations.} Nevertheless, we will argue that the putative EFT is in tension with the bounds from UV/IR mixing discussed in the main text.
For our purposes, we just need the scalar potential \cite{Aparicio:2015psl}
\begin{equation}
    \kappa_4^4 V = \dfrac{a \ab A \ab}{2 \, \e^{a c}} \biggl[ \dfrac{a \ab A \ab}{3c \, \e^{a c}} \Bigl( 1 + \dfrac{3}{a c} \Bigr) - \dfrac{\ab W_0 \ab}{c^2} \biggr] + \dfrac{\sigma^2}{c^2}.
\end{equation}
where $\kappa_4 \varphi = \sqrt{3/2} \, \ln c$ is the canonical Einstein-frame radion, with a stabilized axion.
All other fields are assumed stabilized at higher energy scales and appear with their expectation values inside all terms: the Gukov-Vafa-Witten superpotential $W_0$, the constants $a$ and $A$, depending on the details of non-perturbative correction, and the doubled tension $2 \tau_3 = m_{\p, 4}^4 \sigma^2/c^2$ of an uplifting anti-D3-brane at the tip of a warped throat \cite{Kachru:2003sx}.
For appropriate parameters, the potential has a unique metastable vacuum at large volume $c \gg 1$, such that
\begin{align}
    \ab W_0 \ab & = \dfrac{2 \ab A \ab}{3} \, \Bigl( 1 + \dfrac{3}{2 a c} \Bigr) \dfrac{a c}{\e^{a c}} + \dfrac{4 \sigma^2}{a^2 \ab A \ab} \, \dfrac{\e^{a c}}{\Bigl( 1 + \dfrac{2}{a c} \Bigr) c }.
\end{align}
If the uplift term is such that
\begin{align}
    \sigma^2 = \dfrac{a^2 \ab A \ab^2}{6} \, \Bigl( 1 + \dfrac{2}{a c} \Bigr) \, c \, \e^{-2 a c},
\end{align}
the vacuum is Minkowski.
For a tiny (positive) cosmological constant, we can still take the value of $c$ fixed by the previous relation as a good approximation of the volume vacuum expectation value.

To fulfill eq.~(\ref{varphi - lower bound}), one needs
\begin{align}
    c \geq c_1 = \biggl[ \dfrac{m_{\p, 4}^4}{V_1} \biggr]^{s},
\end{align}
with $s = p/[\sqrt{6} (1-2q) \lambda]$, which is expected to be an order-one parameter.
If $c_1 \ll c$, the observationally-viable potential is dominated by the $\sigma$-independent positive term, and one should solve $\smash{(a^2 \ab A \ab^2/6)^s \, c_1^{1-s} = \e^{2 a s c_1}}$: because of the exponential term, at $c_1 \gg 1$ -- which is necessary for perturbative control -- there is no solution.
If $c_1 \simeq c$, since near the minimum the potential is tiny -- \emph{i.e.} $\smash{\langle V \rangle \simeq 10^{-120} m_{\p, 4}^4}$ --, the power should be very small for the volume $c^{3/2} \gtrsim 10^{180 s}$ to allow for an observationally-viable KK scale, besides being compatible with the microscopic setup itself.
These claims can be supported numerically.
Our conclusion that KKLT violates holographic bounds -- within the controlled EFT region -- is in line with previous results, such as \cite{Conlon:2012tz, Ooguri:2018wrx,Blumenhagen:2022dbo}.

\subsection{LVS}
The Large Volume Scenario (LVS) is another proposal for a dS vacuum in string theory which relaxes the degree of fine tuning of KKLT through a specific internal geometry and with the inclusion of $\alpha'$-corrections \cite{Balasubramanian:2005zx}.

The EFT features two kinetically-mixed moduli $\tau_b$ and $\tau_s \ll \tau_b$, such that the Einstein-frame compactification volume scales as $\smash{\tau_b^{3/2} - \tau_s^{3/2} \simeq \tau_b^{3/2}}$.
Although our discussion does not apply directly to a multi-field model, we can explore the bound along the $\tau_b$-direction, if $\tau_s$ sits at its minimum.
In this case, one can take $\smash{\tau_b = \e^{\sqrt{2/3} \, \kappa_4 \varphi}}$ and the potential takes the form\footnote{In the absence of a suitable uplift, the LVS scalar potential is
\begin{align*}
    \dfrac{V}{m_{\p, 4}^4} = 4 a \ab A \ab \biggl[ \dfrac{2 a \ab A \ab}{3} \, \dfrac{\tau_s^{\frac{1}{2}}}{\tau_b^{\frac{3}{2}}} \, \e^{- a \tau_s} - \ab W_0 \ab \, \dfrac{\tau_s}{\tau_b^3} \biggr] \, \e^{- a \tau_s}
    + \dfrac{3 \xi}{2} \, \dfrac{\ab W_0 \ab^2}{\tau_b^{\frac{9}{2}}},
\end{align*}
where $\xi$ controls the $\alpha'$-corrections.
In the anti-dS vacuum, with $\smash{\langle V \rangle / m_{\p, 4}^4 = -(3/4 a) \, \ab W_0 \ab^2/\langle \tau_s \rangle \langle \tau_b \rangle^{9/2}}$, the solution reads
\begin{align*}
    \langle \tau_b \rangle^{\frac{3}{2}} & = \dfrac{3}{4 a} \, \dfrac{\ab W_0 \ab}{\ab A \ab} \, \langle \tau_s \rangle^{\frac{1}{2}} \, \e^{a \langle \tau_s \rangle}, \\
    \langle \tau_s \rangle & = \xi^{\frac{2}{3}}.
\end{align*}}
\begin{align}
    V = \dfrac{\mu_1^4}{\tau_b^{\frac{3}{2}}} - \dfrac{\mu_2^4}{\tau_b^3} + \dfrac{\mu_3^4}{\tau_b^{\frac{9}{2}}} + \dfrac{\mu_4^4}{\tau_b^{3n}},
\end{align}
for four constants $\mu_{1,2,3,4}$, with the last term representing a generic uplift term.
It is apparent that the tensions that appear in the KKLT potential are evaded: as we are just balancing different powers, it is conceivable to expect solutions to $\smash{\tau_{b,1} = (m_{\p, 4}^4/V_1)^s}$ such that $\smash{1 \ll \tau_{b,1} < \langle \tau_b \rangle}$, where $\langle \tau_b \rangle$ is the standard LVS solution.
In this case, knowledge of all the parameters is crucial.

\bibliographystyle{apsrev4-1}
\bibliography{refs.bib}

\end{document}